\documentclass[%
preprint,superscriptaddress,
amsmath,amssymb,prd,aps,showpacs
floatfix,
]{revtex4-1}
\usepackage{graphicx}
\usepackage{dcolumn}
\usepackage{bm}
\usepackage{mathrsfs}
\usepackage{amsmath}
\usepackage{amssymb}
\usepackage{float}
\usepackage{dcolumn}
\usepackage{multirow}
\usepackage{dsfont}

\begin{document}
\title{Stochastic fluctuations in an eco-evolutionary game dynamics with environmental feedbacks}
\author{Chao Wang \footnote{e-mail: chaowang@nwpu.edu.cn.}}
\affiliation{ School of Ecology and Environment, Northwestern Polytechnical University, Xi'an 710072, China}
\author{Minlan Li \footnote{e-mail: minlanli@mail.nwpu.edu.cn.}}
\affiliation{ School of Ecology and Environment, Northwestern Polytechnical University, Xi'an 710072, China}
\author{Chang Liu \footnote{e-mail: changliu@cc.nara-wu.ac.jp.}}
\affiliation{ Department of Physics, Nara Women's University, Nara, 630-8506, Japan}


\begin{abstract}
Building upon the eco-evolutionary game dynamics framework established by Tilman et al., we investigate stochastic fluctuations in a two-strategy system incorporating environmental feedback mechanisms, where the payoff matrix exhibits population size dependence. We adopt a systematic approach which is the so-called $\Omega$-expansion. When the stochastic factor is integrated, it is shown that the population size for each strategy fluctuates around the interior equilibrium of the macroscopic equations (corresponding to the deterministic model of the eco-evolutionary game) and its variance converges to a constant that is proportional to the environmental carrying capacity if the interior equilibrium is asymptotically stable. The simulation results demonstrate that the $\Omega$ expansion provides a valid approximation, and the reliability of the aforementioned conclusions is verified. Therefore, analogous to Fudenberg and Harris' s stochastic replicator dynamics for infinite populations under external noise (\emph{J. Econ. Theory 57, 420-441}), the dynamic stability of the eco-evolutionary game can be extended to the stochastic regime when the environmental carrying capacity is sufficiently large.
\end{abstract}
\maketitle


\section{Introduction}
Since the 1970s, evolutionary game theory has not only achieved great success in explaining the evolution of animal behavior but has also been widely applied in economics, social sciences, human behavioral ecology, etc \cite{hofbauer1998,nash1950,smith1982,lawlor1976}. The theory provides a fundamental mathematical framework for the evolution of these fields. However, in the traditional evolutionary game theory, environmental changes are neglected in time. As pointed by Tilman et al. \cite{tilman2020} real-world systems often characterized by bi-directional feedbacks between the environment and the incentives in strategic interactions: an individual' s payoff depends not only on her actions relative to the population, but also on the state of the environment, and the state of the environment is influenced by the actions adopted by individuals in the population \cite{tilman2020,chao2023}. Reciprocal feedbacks between strategies and the environment play out in many complex systems, for example, in social-ecological systems, evolutionary-ecological systems, and even psychological-economic systems, et al \cite{Sethi1996,Lacitignola2007,Tavoni2012,Levin2013,Innes2013,Richter2015,tilman2020,Tilman2018,Mullon2018,Estrela2018}.

Recently, Tilman et al. developed a general framework for eco-evolutionary games in which evolutionary game dynamics are environmentally coupled, which is called eco-evolutionary game theory \cite{tilman2020}. They investigated the stability of the eco-evolutionary game dynamics involving the time evolution of both strategies and environmental factors and showed how the joint dynamics of strategies and the environment depend on the fitness of individuals and the relative speed of environmental versus strategic change (see also Ref.~\cite{Grunert2021}). Following eco-evolutionary game theory, we studied the evolutionary stability and the ecologically evolutionarily stale strategy (eco-ESS) in the eco-evolutionary game dynamics with density dependence \cite{chao2023,yan2024,Minlan2025}. We also found that there exists a critical value of which the cooperators emerging spatial structured population when the environmental feedback is taken into account \cite{yan2024}.

On the other hand, fluctuations are predigested in the framework introduced by Maynard and Price \cite{Peck1988,hofbauer1998,lawlor1976,smith1982}. Traditional studies on evolutionary game theory are based on deterministic differential equations that assume an infinite population. When the population size is finite, demographic fluctuation-intrinsic noise has implications for the system \cite{foster1990,hastings2004}. In this case, the dynamic is described by the stochastic process theory and characterized by the concept of the fixation probability for a fixed population size \cite{nowak2004}. Then, the stochastic evolutionary dynamics of the system remain a critical subject for investigation when the population size is not fixed \cite{yi2007,anna2010}. In particular, when the exogenous environment changes, it is difficult to guarantee the population size to be infinite \cite{chao2023,chen2018,anna2010}. It is worth studying the stochastic fluctuations of eco-evolutionary games when environmental feedbacks are considered.

In the present study, we investigate stochastic fluctuations and the local stability of the equilibrium in a two-strategy eco-evolutionary game involving the environmental feedbacks, where the payoff matrix is density-dependent. Within the assumption of a large environmental carrying capacity, we introduce a systematic approach which is the so-called $\Omega$ expansion method and widely used in stochastic processes \cite{Kampen1992}. Based on this method, the population dynamic can be divided into two parts: the macroscopic equations and the moment equations. The macroscopic equations are equivalent to the deterministic model. The feature of stochastic fluctuations is described by the moment equations. Similar to the stochastic replicator dynamics with infinite populations subject to external noise conducted by Fudenberg and Harris \cite{Fudenberg1992,traulsen2006}, we found that the population fluctuates around the interior state and the fluctuation range is constant if the interior state is asymptotically stable in the eco-evolutionary game. Unlike the intrinsic noise of the evolutionary game dynamics \cite{Fudenberg1992}, the fluctuation is related to the environmental carrying capacity instead of the population size in the eco-evolutionary game dynamics. We also performer a stochastic simulation based on Gillespie's algorithm as a confirmation of the theoretical derivation results.

The remainder of this paper is organized as follows. In Sect.~II, we provide the basic formula with the deterministic model and stochastic model based on a two-strategy eco-evolutionary game with environmental feedbacks, where the population size is density dependence. In Sect.~III, the expansion is performed by the environmental carrying capacity (denoted by $\Omega$ for simplicity). In Sect.~IV, we study the stochastic fluctuations near the steady-state and investigate the local stability of the equilibrium. In the Sect.~V, a simulation independent with analytic derivation is performed to verify aforementioned results. Finally, a summary is given in the last section.

\section{The basic model}

We begin by considering a two-strategy (denoted as $A$ and $B$) evolutionary game. The number of individuals adopting strategy $A(B)$ in the group is $n_{1(2)}$. And the population size is then denoted as $N(=n_1+n_2)$. The payoff matrix of the two strategies depends on the $N$ and is given by the payoff matrix \cite{chao2023}
\begin{eqnarray}
\begin{array}{cc}
   \begin{array}{c}
    \\
    \begin{array}{c}
    A\\B
    \end{array}
  \end{array}&

  \begin{array}{c}
    \begin{array}{cc}
    A & \qquad B
    \end{array}
    \\
    \left(\begin{array}{cc}
    \alpha_{11}(N)& \alpha_{12}(N)\\
    \alpha_{21}(N)& \alpha_{22}(N)
    \end{array} \right).
  \end{array}
\end{array}
\end{eqnarray}
Moreover, we also assume that $\mathrm{d}\alpha_{ij}(N)/\mathrm{d}N<0$, that is each entry of the payoff matrix is a decreasing function $N$. Following our previous work, we consider $\alpha_{ij}(N)=a_{ij}-b_{ij}N$, where both $a_{ij}$ and $b_{ij}$ are constants with $b_{ij}>0$. In a well-mixed population, individuals interact in random pairwise contests.

In the model used in this study, the population size ($N$) was not fixed. To include the population size effects, the fitness for each strategy is given by $W=W_0+f$, where $W_0$ is the background fitness \cite{smith1982,yi2007}. That is, $W_0$ is the component of an individual's fitness due to changes in the environments \cite{smith1982,tilman2020,yi2007,chao2023}. Based on payoff matrix (1), the corresponding fitness can be given as follows \cite{yi2007,smith1982,tilman2020}
\begin{eqnarray}
W_i=W_{0i}+f_i=1-(x_1b_{i1}N+x_2b_{i2}N)+\omega( x_1a_{i1}+ x_1a_{i2}),
\end{eqnarray}
where $W_{0i}=1-(x_1b_{i1}N+x_2b_{i2}N)$ and $f_i=\omega( x_1a_{i1}+ x_1a_{i2})$ with $\omega$ being the selection intensity \cite{nowak2004}, $x_{1(2)}[=\frac{n_{1(2)}}{n_1+n_2}]$ denotes the frequency of the individual taking the strategy $A(B)$. Without a loss of generality, we set $b_{11}=b_{12}=b_{21}=b_{22}=b$, thereby reducing $W_0$ to $1-bN$. The positive parameter $1/b$ can be thought of as an environmental carrying capacity $\Omega$.

To obtain a more general result, we treat the birth rate as a non-linear function of expected payoffs, $u_i=\exp [W_i]$ \cite{chao2023}. As in our previous work, all individuals have the same death rate, denoted by $\beta$ with $0<\beta<1$ \cite{chao2023}. The population growth dynamics can then be described by the following equations
\begin{eqnarray}
\frac{\mathrm{d}n_i}{\mathrm{d}t}&=&u_i n_i-\beta n_i.
\end{eqnarray}
We assume that the selection intensity ($\omega$) is given by the utilization rate of the environmental carrying capacity, i.e., $\omega=N/\Omega$ and define an expanded frequency of $p_i=n_i/\Omega$, Eqs.~(3) can be rewritten as
\begin{eqnarray}
\frac{\mathrm{d}p_1}{\mathrm{d}t}&=& p_1 [\mathrm{e}^{1-(p_1+p_2)+e_1}-\beta],\nonumber\\[8pt]
\frac{\mathrm{d}p_2}{\mathrm{d}t}&=& p_2 [\mathrm{e}^{1-(p_1+p_2)+e_2}-\beta],
\end{eqnarray}
with $e_1=a_{11} p_1+a_{12} p_2$ and $e_2=a_{21} p_1+a_{22} p_2$. In this formulation, environmental feedbacks are explicitly incorporated into the fitness expression. Not only the payoffs generated by playing games are the source of individual incomes, but also an environmental factors (population size effects) are valid in the birth rate function. We call this equation eco-evolutionary game dynamics with density dependence as Refs. \cite{chao2023,tilman2020}. If the population size is assumed to be sufficiently large, Eqs.~(4) can be replaced by the replication equation with the fitness given by $u_i-\beta$.

Next, we expand this deterministic model to the stochastic case by approximating the one step process, in which the intrinsic noise is taken into account \cite{swift2002,Nasell2001,yi2007}. We introduce a joint probability distribution $\Phi(n_1,\,n_2;\,t)$ which corresponds to the state of the numbers of the strategies $A$ and $B$ being $n_1$ and $n_2$, respectively, at time $t$. Following the so-called ``birth-and-death processes" in van Kanmpen's book \cite{Kampen1992}, we assume that at most one birth or death event occurs during a sufficiently small time interval. At each time, each individual has a death rate $\beta_1=\beta n_1$ (for strategy $A$) or $\beta_2=\beta n_2$ (for strategy $B$) and a probability $\alpha_1(n_1,\,n_2) =u_1(n_1,\,n_2) n_1$ (for strategy $A$) or $\alpha_2(n_1,\,n_2)=u_2(n_1,\,n_2) n_2$ (the strategy $B$) per unit time to produce a second one by replication. Subsequently, the joint probability distribution $\Phi(n_1,\,n_2;\,t)$ jumps with the following probabilities \cite{yi2007,swift2002,Nasell2001}
\begin{eqnarray}
\Phi(n_1,\,n_2;\,t)&\xrightarrow{n_1u_1} &\Phi(n_1+1,\,n_2;\,t+\Delta t),\quad
\Phi(n_1,\,n_2;\,t) \xrightarrow{n_1\beta} \Phi(n_1-1,\,n_2;\,t+\Delta t),\nonumber\\[4pt]
\Phi(n_1,\,n_2;\,t)&\xrightarrow{n_2 u_2} &\Phi(n_1,\,n_2+1;\,t+\Delta t),\quad \Phi(n_1,\,n_2;\,t)\xrightarrow{n_2\beta} \Phi(n_1,\,n_2-1;\,t+\Delta t).
\end{eqnarray}
We emphasize that this one-step process is distinct from the Moran process due to its non-conservation of the total population size.

The master equation of $\Phi(n_1,\,n_2;\,t)$ can thus be given by
\begin{eqnarray}
\frac{\partial \Phi(n_1,\,n_2;\,t)}{\partial t}=\sum^2_{i=1}[(\mathds{E}^-_i -1) u_i(n_1,\,n_2) n_i \phi(n_1,\,n_2;\,t) +(\mathds{E}^+_i -1)\beta n_i \phi(n_1,\,n_2;\,t)],\nonumber\\
\end{eqnarray}
where the symbol $\mathds{E}_i^\pm$ represents a step operator, defined by its effect on an arbitrary function $f(n)$: $\mathds{E}_i^\pm f(n_i)=f(n_i\pm1)$. Considering the continuity requirements of $\Phi(n_1,\,n_2;\,t)$, the operator $\mathds{E}_i^\pm$ could be approximated as a Taylor expansion \cite{Kampen1992}
\begin{eqnarray}
\mathds{E}_i^\pm =1\pm \frac{\partial}{\partial n_i}+\frac{1}{2}\frac{\partial^2}{\partial n_i^2}\pm\frac{1}{3!}\frac{\partial^3}{\partial n_i^3}+\cdots.
\end{eqnarray}
Omitting orders higher than the second, we obtain the Fokker-Planck equation for the master equation
\begin{eqnarray}
\frac{\partial \Phi(n_1,\,n_2;\,t)}{\partial t}&=&\sum^2_{i=1}\left[-\frac{\partial}{\partial n_i} \left(\alpha_i-\beta_i\right) \phi(n_1,\,n_2;\,t) +\frac{1}{2}\frac{\partial^2}{\partial n_i^2}  \left(\alpha_i +\beta_i \right)\phi(n_1,\,n_2;\,t) \right].
\end{eqnarray}
This is a non-linear Fokker-Planck equation with the coefficients including $n_1$ and $n_2$. Therefore, it is impossible to obtain the exact solution. In the next section, we will study its stochastic behavior using the technique of a power series expansion in the parameter $\Omega$ developed by van Kanmpen \cite{Kampen1992}.

\section{The $\Omega$ expansion}
To anticipate the way in which the solution $\Phi(n_1,\,n_2;\,t)$ will depend on $\Omega$, one expects that $\Phi(n_1,\,n_2;\,t)$ is a sharp peak around the macroscopic values $n_{1(2)}=\Omega p_{1(2)}$ while its width will be of order $n_{1(2)}^{-1/2}\thicksim\Omega^{-1/2}$ for intrinsic noise \cite{Kampen1992}. Therefore, we assume the following
\begin{eqnarray}
n_1=\Omega p_1+\Omega^{1/2} \xi, \qquad n_2=\Omega p_2+\Omega^{1/2} \eta,
\end{eqnarray}
where the first terms in each equation are macroscopic quantities, and $\xi$ and $\eta$ are two new variables that replace $n_{1} $ and $n_2$, respectively. Then, the joint probability distribution $\Phi(n_1,\,n_2;\,t)$ can be written as a function of $\xi$ and $\eta$, $\Pi(\xi,\,\eta;\,t)=\Phi(n_1,\,n_2;\,t)$. We found that $\Pi(\xi,\,\eta;\,t)$ satisfies the linear Fokker-Planck equation in the order of $\Omega^0$ and becomes a two-dimensional Gaussian distribution at each time point. Therefore, It is sufficient to determine the first and second moments of $\eta$ and $\xi$. This framework is called the linear noise approximation.

According to Eqs.~(9), $n_1\rightarrow n_1\pm 1$ and $n_2\rightarrow n_2\pm 1$ are equivalent to $\xi \rightarrow \xi \pm \Omega^{-1/2}$ and $\eta \rightarrow \eta \pm \Omega^{-1/2} $, respectively. Then, the step operator can be expressed as \cite{Kampen1992}
\begin{eqnarray}
\mathds{E}_1^\pm&=&1 \pm \Omega^{-1/2}\frac{\partial}{\partial \xi} +\frac{1}{2}\Omega^{-1}\frac{\partial^2 }{\partial \xi^2 }+\cdots,\nonumber\\[6pt]
\mathds{E}_2^\pm&=&1 \pm \Omega^{-1/2}\frac{\partial}{\partial \eta} +\frac{1}{2}\Omega^{-1}\frac{\partial^2 }{\partial \eta^2 }+\cdots.
\end{eqnarray}
Substituting Eqs.~(9) and (10) into Eqs.~(8), the mater equation in the new variable takes the form
\begin{eqnarray}
\quad\frac{\partial \Phi(n_1,\,n_2;\,t)}{\partial t}
&=&\frac{\partial \Pi(\xi,\,\eta;\,t)}{\partial t}-\Omega^{1/2} \frac{\mathrm{d} p_1(t) }{\mathrm{d}t}\frac{\partial \Pi(\xi,\,\eta;\,t)}{\partial \xi}- \Omega^{1/2} \frac{\mathrm{d} p_2(t)}{\mathrm{d}t}\frac{\partial \Pi(\xi,\,\eta;\,t)}{\partial \eta}\nonumber\\[8pt]
&=& \Omega^{-1/2}\frac{\partial }{\partial \xi}\left[\beta-\mathrm{e}^{1-(p_1+p_2)+e_1-\Omega^{-1/2}(\xi+\eta)+\Omega^{-1/2}(\xi a_{11}+\eta a_{12})}\right](\Omega p_1+\Omega^{1/2} \xi)\Pi(\xi,\,\eta;\,t) \nonumber\\[8pt]
&&+\frac{1}{2}\Omega^{-1}\frac{\partial^2}{\partial \xi^2} \left[\beta+\mathrm{e}^{1-(p_1+p_2)+e_1-\Omega^{-1/2}(\xi+\eta)+\Omega^{-1/2}(\xi a_{11}+\eta a_{12})}\right](\Omega p_1+\Omega^{1/2} \xi)\Pi(\xi,\,\eta;\,t) \nonumber\\[8pt]
&&+ \Omega^{-1/2}\frac{\partial }{\partial \eta}\left[\beta-\mathrm{e}^{1-(p_1+p_2)+e_2-\Omega^{-1/2}(\xi+\eta)+\Omega^{-1/2}(\xi a_{21}+\eta a_{22})}\right](\Omega p_2+\Omega^{1/2} \eta)\Pi(\xi,\,\eta;\,t) \nonumber\\[8pt]
&&+\frac{1}{2} \Omega^{-1}\frac{\partial^2}{\partial \eta^2} \left[\beta+\mathrm{e}^{1-(p_1+p_2)+e_2-\Omega^{-1/2}(\xi+\eta)+\Omega^{-1/2}(\xi a_{21}+\eta a_{22})}\right](\Omega p_2+\Omega^{1/2} \eta)\Pi(\xi,\,\eta;\,t).\nonumber\\
\end{eqnarray}
We reorganized this equation and collected terms for each order of $\Omega$. Equations of different orders correspond to different evolution time scales. These terms can be discussed separately and safely. The following equations were then obtained:
\begin{eqnarray}
 \frac{\mathrm{d} p_1(t) }{\mathrm{d}t}\frac{\partial \Pi(\xi,\,\eta;\,t)}{\partial \xi} &=& \frac{\partial }{\partial \xi}\left[\mathrm{e}^{1-(p_1+p_2)+e_1}-\beta\right] p_1 \Pi(\xi,\,\eta;\,t),\nonumber\\[8pt]
 \frac{\mathrm{d} p_2(t) }{\mathrm{d}t}\frac{\partial \Pi(\xi,\,\eta;\,t)}{\partial \eta} &=& \frac{\partial }{\partial \eta}\left[\mathrm{e}^{1-(p_1+p_2)+e_2}-\beta\right] p_2 \Pi(\xi,\,\eta;\,t),
\end{eqnarray}
for the $\Omega^{1/2}$ order, and
\begin{eqnarray}
\frac{\partial \Pi(\xi,\,\eta;\,t)}{\partial t}&=& \frac{\partial }{\partial \xi}\left[\left(\beta-\mathrm{e}^{1-(p_1+p_2)+e_1}\right)\xi+ p_1 \mathrm{e}^{1-(p_1+p_2)+e_1}((a_{11}-1)\xi+(a_{12}-1)\eta)\right]\Pi(\xi,\,\eta;\,t) \nonumber\\[8pt]
&&\quad+ \frac{\partial }{\partial \eta}\left[\left(\beta-\mathrm{e}^{1-(p_1+p_2)+e_2}\right)\eta+ p_2 \mathrm{e}^{1-(p_1+p_2)+e_2}((a_{21}-1)\xi+(a_{22}-1)\eta)\right]\Pi(\xi,\,\eta;\,t) \nonumber\\[8pt]
&&\quad+\frac{1}{2}\left[\frac{\partial^2}{\partial \xi^2}\left[\beta+\mathrm{e}^{1-(p_1+p_2)+e_1}\right]p_1+\frac{\partial^2}{\partial \eta^2}\left[\beta+\mathrm{e}^{1-(p_1+p_2)+e_2}\right]p_2\right]\Pi(\xi,\,\eta;\,t),\nonumber\\
\end{eqnarray}
for the $\Omega^0$ order, whereas those with higher orders is neglected.

For the terms of the leading order $\Omega^{\frac{1}{2}}$, the macroscopic equations are established by the coefficients of Eqs. (12),
\begin{eqnarray}
 \frac{\mathrm{d} p_1(t) }{\mathrm{d}t}&=&p_1  \left[\mathrm{e}^{1-(p_1+p_2)+e_1}-\beta\right] ,\nonumber\\[8pt]
 \frac{\mathrm{d} p_2(t) }{\mathrm{d}t}&=&p_2  \left[\mathrm{e}^{1-(p_1+p_2)+e_2}-\beta\right] ,
\end{eqnarray}
which are exactly the two deterministic equations (4). They can be solved explicitly and the globally stable can be obtained if the deterministic equations are solvable. The Eqs.~(13) for the order of $\Omega^0$ is a multivariate linear Fokker-Plank equation. The moments of $\xi$ and $\eta$ are obtained as
\begin{eqnarray}
\partial_t\langle\xi\rangle &=&\left(\beta-\mathrm{e}^{1-(p_1+p_2)+e_1}\right)\langle\xi\rangle + p_1 \mathrm{e}^{1-(p_1+p_2)+e_1}\left[(a_{11}-1)\langle\xi\rangle +(a_{12}-1) \langle\eta\rangle\right],\nonumber\\[4pt]
\partial_t\langle\eta\rangle &=&\left(\beta-\mathrm{e}^{1-(p_1+p_2)+e_2}\right)\langle\eta\rangle + p_2 \mathrm{e}^{1-(p_1+p_2)+e_2}\left[(a_{21}-1)\langle\xi\rangle +(a_{22}-1) \langle\eta\rangle\right].
\end{eqnarray}
On the second order, there are three moments and they obey coupled equations
\begin{eqnarray}
\partial_t\langle\xi^2\rangle &=& 2 \left[\beta-\mathrm{e}^{1-(p_1+p_2)+e_1}+p_1 \mathrm{e}^{1-(p_1+p_2)+e_1} (a_{11}-1)\right]\langle\xi^2\rangle \nonumber\\[4pt]
&&\qquad+2p_2\mathrm{e}^{1-(p_1+p_2)+e_2}(a_{21}-1)  \langle\xi\eta\rangle+\left[\beta+\mathrm{e}^{1-(p_1+p_2)+e_2}\right]p_2,\nonumber\\[4pt]
\partial_t\langle\xi\eta\rangle &=&\Large[2 \beta-\mathrm{e}^{1-(p_1+p_2)+e_1}-\mathrm{e}^{1-(p_1+p_2)+e_2}+p_1 \mathrm{e}^{1-(p_1+p_2)+e_1} (a_{11}-1)\nonumber\\[4pt]
&&\qquad +p_2 \mathrm{e}^{1-(p_1+p_2)+e_2} (a_{22}-1)\Large]\langle\xi\eta\rangle +2p_2\mathrm{e}^{1-(p_1+p_2)+e_2}(a_{21}-1)  \langle\xi^2\rangle\nonumber\\[4pt]
&&\qquad\qquad +2p_1\mathrm{e}^{1-(p_1+p_2)+e_1}(a_{12}-1) \langle\eta^2 \rangle, \nonumber\\[4pt]
&&\qquad+2p_1\mathrm{e}^{1-(p_1+p_2)+e_1}(a_{12}-1)  \langle\xi\eta\rangle+\left[\beta+\mathrm{e}^{1-(p_1+p_2)+e_1}\right]p_1,\nonumber\\[4pt]
\partial_t\langle\eta^2\rangle &=& 2 \left[\beta-\mathrm{e}^{1-(p_1+p_2)+e_2}+p_2 \mathrm{e}^{1-(p_1+p_2)+e_2} (a_{22}-1)\right]\langle\eta^2\rangle\nonumber\\[4pt]
&&\qquad+2p_1\mathrm{e}^{1-(p_1+p_2)+e_1}(a_{12}-1)  \langle\xi\eta\rangle+\left[\beta+\mathrm{e}^{1-(p_1+p_2)+e_1}\right]p_1,
\end{eqnarray}
where the exponential decay function can be further expanded and simplified in the power of $\Omega$. The form is given in Appendix A. From Eqs.~(15) and (16), one can determine the variances and the covariance of the fluctuations of $n_i$ around the macroscopic solution of Eqs.~(14). If we define the initial distribution as a monotropic function $\Phi(n_1,\,n_2;\,0)=\delta(n_{1}-n_{10})\delta(n_{2}-n_{20})$, and the solutions of the macroscopic equations as $p_i(t,\,p_{i0})$ for the initial condition $p_i(0)=p_{i0}$, the variances and covariance of $n_i$ at the time $t$ based on the definition of Eqs.~(8) is therefore determined by
\begin{eqnarray}
\langle n_1(t)\rangle &=& \Omega p_1(t,\,p_{10})+ \Omega^{1/2} \langle\xi(t)\rangle,\quad \langle \langle n_1^2(t) \rangle \rangle = \Omega \langle\langle\xi^2(t)\rangle\rangle,\nonumber\\
\langle n_2(t)\rangle &=& \Omega p_2(t,\,p_{20})+ \Omega^{1/2} \langle\xi(t)\rangle,\quad \langle \langle n_2^2(t) \rangle \rangle = \Omega \langle\langle\eta^2(t)\rangle\rangle.
\end{eqnarray}
where are the moments related to $\xi$ and $\eta$ can be solved from Eqs.~(15) and (16), and the symbol $\langle \langle - \rangle \rangle$ represents the variance of the variable. We can see that the features of stochastic fluctuations are determined by the macroscopic equations (being equivalent to the deterministic model (4) ) and the moments of $\eta$ and $\xi$.

\section{STOCHASTIC FLUCTUATIONS NEAR THE STEADY-STATE}
Recently, to develop the concept of evolutionary stability in the stochastic model, Zheng et al. \cite{zheng2017,zheng2018} investigated the conditions for stochastic local stability of the fixation states and constant interior equilibria in a two-phenotype model with random payoffs. According to our previous work, the deterministic model in Eqs.~(4) has at most two interior equilibria. In the case of the stochastic model, to investigate the local stability of the equilibria, we concentrate on the dynamics in the neighborhood of the equilibria \cite{zheng2017,zheng2018}. Therefore, without loss of generality, we specialize in systems possessing one interior equilibrium. In the deterministic population, the interior equilibrium is stable if $a_{12}>a_{22}$ and $a_{11}<a_{21}$ \cite{chao2023}. We will extend this to a stochastic model based on Eqs.~(14)-(16).

\subsection{Dynamical stability of the macroscopic equation}
The macroscopic equations (14) is equivalent to the deterministic equations (4). This reveals the mean value of the growth equations for the population in such a way that the fluctuations around it are of the order $\Omega^{1/2}$. We had discussed the deterministic equations in detail before \cite{chao2023}. In the eco-evolutionary game, it was recorded that macroscopic equations Eqs.~(14) has one interior equilibrium:
\begin{eqnarray}
p_1^*&=&\frac{(\ln \beta-1)(a_{12}-a_{22})}{a_{11}a_{22}-a_{12}a_{21}-a_{11}-a_{22}+a_{12}+a_{21} },\nonumber\\[8pt]
 p_2^*&=&\frac{(\ln \beta-1)(a_{21} -a_{11} )}{a_{11}a_{22}-a_{12}a_{21}-a_{11}-a_{22}+a_{12}+a_{21} },
\end{eqnarray}
where $a_{11}a_{22}-a_{12}a_{21}-a_{11}-a_{22}+a_{12}+a_{21} <0$. The Jacobi matrix about this interior equilibrium are
\begin{eqnarray}
  \left ( \begin{array}{ccc}
  p_1^*\beta( a_{11}-1) & \qquad p_1^*\beta (a_{12}-1)  \\
  p_2^*\beta (a_{21}-1) & \qquad p_2^*\beta (a_{22}-1) \\
  \end{array}\right).
\end{eqnarray}
Note that the characteristic equation of this matrix is
\begin{eqnarray}
\lambda^2-\beta[ p_1^*(a_{11}-1)+p_2^* (a_{22}-1) ]\lambda-\beta^2 p_1^* p_2^*  [(a_{12}-1) (a_{21}-1)-( a_{11}-1) (a_{22}-1)]=0,\nonumber\\
\end{eqnarray}
It is easily to obtained that $(p_1^*,\,p_2^*)$ is asymptotically stable \cite{chao2023}. This result is similar to that of the classic two-phenotype replicator equation.

\subsection{THE STOCHASTIC LOCAL STABILITY OF AN EQUILIBRIUM}
In this study, we concentrate on the local fluctuating properties near the stationary point of the macroscopic equations for the eco-evolutionary game. Within a delta distribution of the initial state, the initial fluctuations vanish $\langle\xi\rangle_0=\langle\eta\rangle_0=\langle\xi^2\rangle_0=\langle\eta^2\rangle_0=\langle\xi\eta\rangle_0=\langle\langle\xi^2\rangle\rangle_0=\langle\langle\eta^2\rangle\rangle_0=\langle\langle\xi\eta\rangle\rangle_0=0$. Substituting the stationary point $(p_1^*,\,p_2^*)$ into Eqs.~(15) and omitting the higher orders through Appendix A, we have
\begin{eqnarray}
\partial_t\langle\xi\rangle^s&=& p_1^*\beta (a_{11}-1) \langle\xi\rangle^s +p_1^*\beta  (a_{12}-1) \langle\eta\rangle^s,\nonumber\\[4pt]
\partial_t\langle\eta\rangle^s &= &p_2^*\beta (a_{21}-1) \langle\xi\rangle^s+p_2^*\beta (a_{22}-1)\langle\eta\rangle^s,
\end{eqnarray}
and Eqs.~(16) are simplified to
\begin{eqnarray}
\partial_t\langle\xi^2\rangle^s &=& 2 \beta p_1^*  (a_{11}-1)  \langle\xi^2\rangle^s +2\beta p_1^* (a_{12}-1) \langle\xi\eta\rangle^s+ 2 \beta p_1^*,\nonumber\\[4pt]
\partial_t\langle\xi\eta\rangle^s &=& 2 \beta p_2^* (a_{21}-1)  \langle\xi^2\rangle^s+\beta \left [   p_1^*  (a_{11}-1)+   p_2^*(a_{22}-1) \right ] \langle\xi\eta\rangle^s + 2\beta p_1^* (a_{12}-1) \langle\eta^2 \rangle^s,\nonumber\\[4pt]
\partial_t\langle\eta^2\rangle^s &=& 2\beta p_2^* (a_{21}-1)   \langle\xi\eta\rangle^s+2\beta p_2^*(a_{22}-1)  \langle\eta^2\rangle^s+2 \beta p_2^*,
\end{eqnarray}
where the superscript $s$ represents the results near the range of the stationary point in the macroscopic solution.

The solutions of $\langle\xi\rangle^s$ and $\langle\eta\rangle^s$ in Eqs.~(21) are presented in Appendix B. The coefficient matrix was confirmed to consistent with the Jacobi matrix of the macroscopic equation. Since the interior equilibrium of the the deterministic equations is asymptotically stable, $(\langle\xi\rangle^s=0,\,\langle\eta\rangle^s=0)$ is also asymptotically stable, which means that $\langle\xi\rangle^s=0$ and $\langle\eta\rangle^s=0$ for $t\rightarrow \infty$. According to Eqs. (16), the averages of $n_1$ and $n_2$ are  therefore completely determined by the macroscopic equation, which is consistent with the deterministic model,  i.e., $\langle n_{1(2)}\rangle_{t\rightarrow \infty} = \Omega p_1^*$.

As for Eqs.~(22) of the variance, which are non-homogeneous linear ordinary differential equation systems, the solution is given in Appendix C. The two eigenvalues of the coefficient matrix are the same as those for Eqs.~(21), and the other eigenvalue is $\lambda_3=2\beta \left [   p_1^*  (a_{11}-1)+   p_2^*(a_{22}-1) \right ]$. According to the Routh-Hurwitz conditions, the variance converges to fixed values. In other words, using Eqs.~(17), the variance of the size for each strategy, $\langle \langle n_{1(2})^2(t) \rangle \rangle$, is a constant for $t\rightarrow \infty$, which is directly proportional to the environmental carrying capacity $\Omega$. We will verify these results in the following simulations.

\section{Numerical results and Discussion}
In this section, we will display a numerical result obtained through simulation and discuss the steady-state statistics for the system near the stationary point of the macroscopic equations. Following the stochastic simulation of coupled chemical reactions by Gillespie \cite{daniel1976,daniel1977,daniel2000}, we implemented a Monte Carlo simulation to account for fluctuations. In this stochastic simulation algorithm, we defined two ``chemically active species" $A$ and $B$ with corresponding sizes being $n_1$ and $n_2$, respectively. Their dynamics are governed by the update rule in Eqs. (5). We will demonstrate that our derived results are consistent with the simulated results. As the simulations started from Eqs.~(5) but not the linear noise approximation in Eqs.~(9), one assures that $\Omega$ expansion will be a good approximation in our formula when the population size is far from the environmental carrying capacity, i.e., $p_i=n_i/\Omega<<1$.

As a representative example, the parameters in our simulation are the payoff matrix $\left ( \begin{array}{ccc}
  -10 & 1  \\
  1 & -10 \\
  \end{array}\right)$, which is in the case of co-coexistence at the interior equilibrium (globally stable) for the deterministic equations, and the death rate $\beta=0.1$. In Fig.~1, we give the results of $10^7$ Monte Carlo steps with $\Omega=10000$. The initial condition is $(n_1,\,n_2)=(\Omega p_1^*,\Omega \,p_2^*)$. It is shown that the population size for each strategy fluctuates around the interior equilibrium of the macroscopic equations and its variance is fixed for a long time, which is consistent with the theoretical result.
\begin{figure}[bt]
\centering
\includegraphics[bb=7 264 643 643, width=0.8\textwidth]{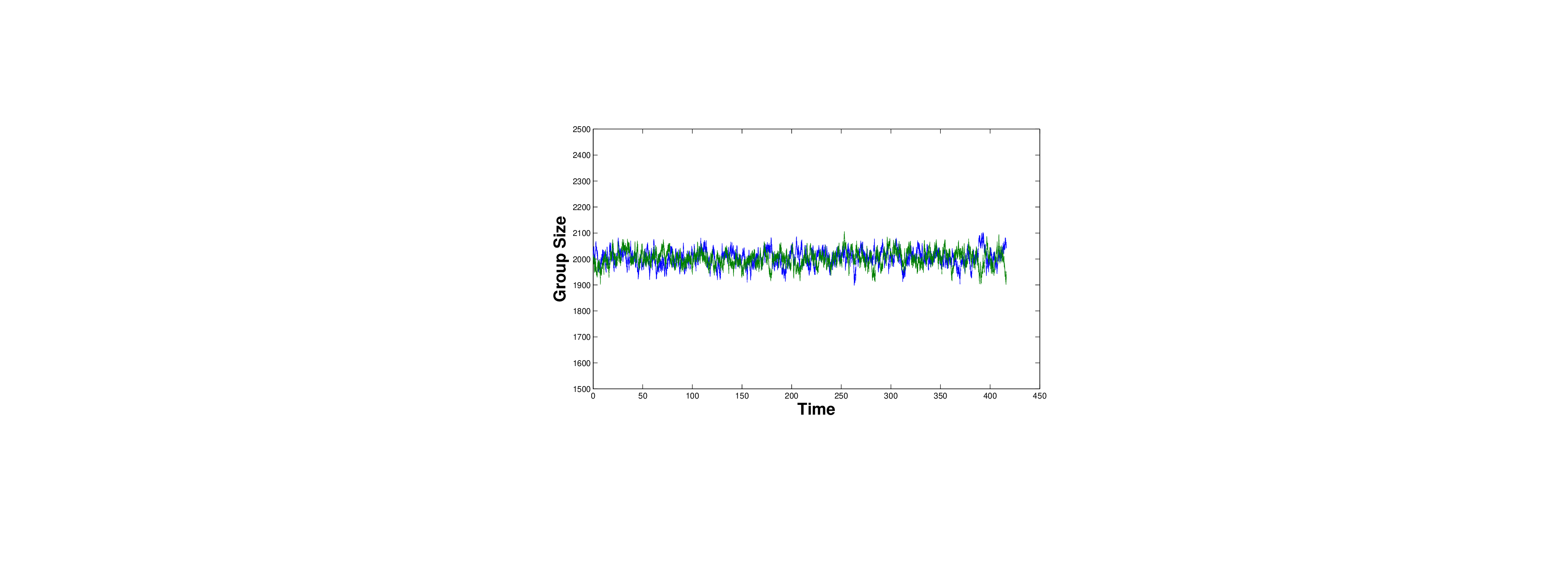}
\caption{Stochastic simulations of the population size $n_1$ and $n_2$ for $10^7$ Monte Carlo step with $\Omega=10000$ starting from the interior equilibrium. The parameters in the simulation are the payoff matrix $\left ( \begin{array}{ccc}
  -10 & 1  \\
  1 & -10 \\
  \end{array}\right)$ and the death rate $\beta=0.1$. The blue and green solid lines correspond to $n_1$ and $n_2$, respectively. }\label{a-f_mixing}
\end{figure}

According to Eqs.~(17), the variance of the fluctuations of $n_i$ is proportional to the environmental carrying capacity $\Omega$. We calculate the variance of $n_i$ in a complete simulation starting from the initial condition $(n_1,\,n_2)=(\Omega p_1^*, \, \Omega p_2^*)$. For each value of $\Omega$, we conducted 500 simulations and took the average. The covariance of $n_i$ with respect to $\Omega$ is displayed in Fig.~2. It is shown that the covariance depends linearly on $\Omega$ and the coefficient is determined by the payoff matrix parameters and the death rate $\beta$. This tendency is consistent with theoretical predictions. The simulation experiment results validated Eqs.~(21) and (22).
\begin{figure}[bt]
\centering
\includegraphics[bb=86 281 595 585, width=0.8\textwidth]{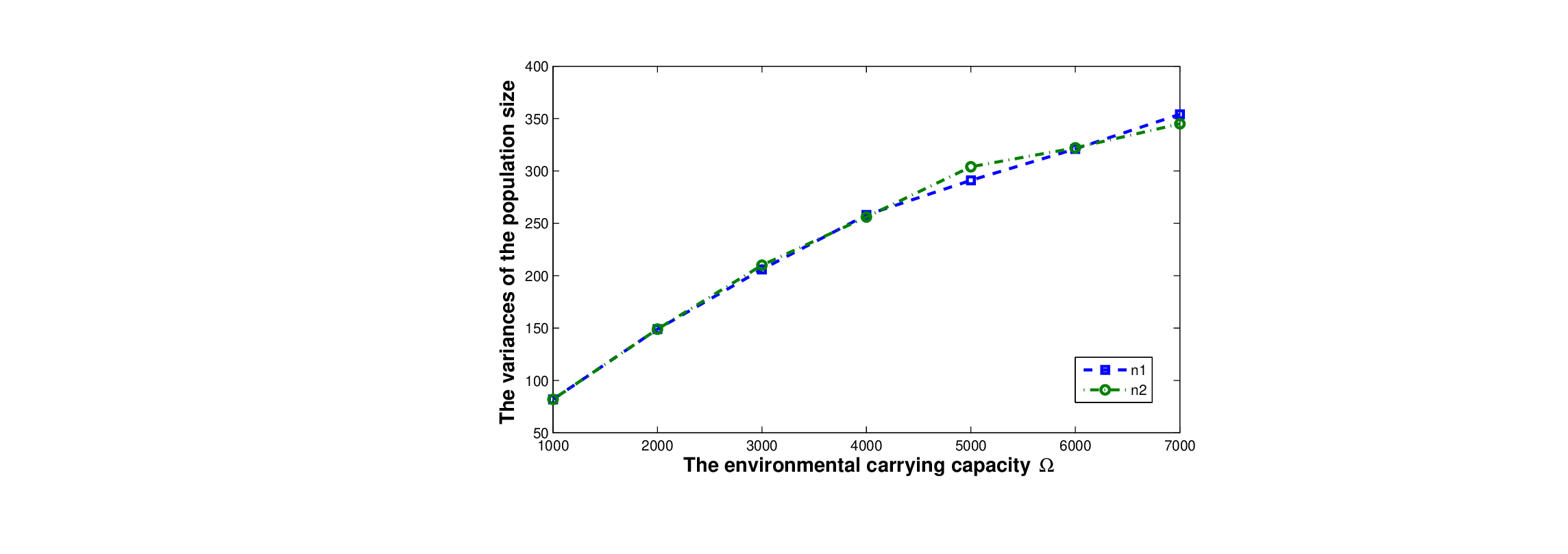}
\caption{The variance of $n_1$ and $n_2$ near the interior equilibrium with respect to $\Omega$. It is shown that the variance depends linearly on $\Omega$ approximately. In each value of $\Omega$, we conducted 500 simulations and took the average. Each simulations is preformed by $10^7$ Monte Carlo step. The parameters in the simulation are the payoff matrix $\left ( \begin{array}{ccc}
  -10 & 1  \\
  1 & -10 \\
  \end{array}\right)$ and the death rate $\beta=0.1$. The blue dashed and green dot-dashed lines correspond to $n_1$ and $n_2$, respectively.}
\end{figure}

In fact, when $n_{1(2)}$ is far from the stationary point, the population size evolves along the direction determined by the deterministic equations. We display the evolutionary trends (Fig.~3) when the initial conditions $p_{1(2)}=0.4(0.1),\, 0.6(0.1)$, respectively, and both processes with different initial conditions converge to the stationary point. However, since the formula of the the moment equations (21) and (22) couples with the macroscopic equations (14) when the high order of $\Omega^{-\frac{1}{2}}$ is taken into account, it is difficult to give quantitative results at present. We will investigate this subject in the future.

\begin{figure}[bt]
\centering
\includegraphics[bb=0 273 621 591, width=\textwidth]{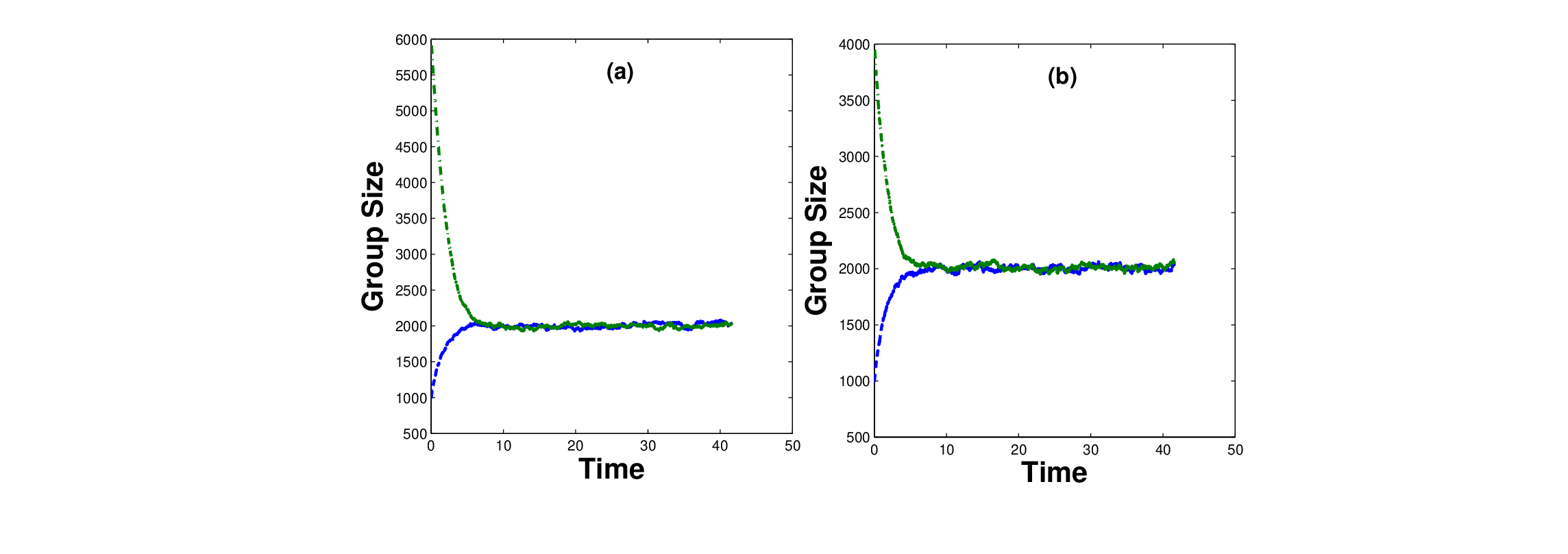}
\caption{Stochastic simulations of the population size with the initial states far from the the stationary point. (a) and (b) correspond to the initial states $n_2=4000\,(n_1=1000)$ and $n_2=6000 \,(n_1=1000)$, respectively. The parameters in the simulation are the payoff matrix $\left ( \begin{array}{ccc}
  -10 & 1  \\
  1 & -10 \\
  \end{array}\right)$ and the death rate $\beta=0.1$. The blue dashed and green dot-dashed solid lines correspond to $n_1$ and $n_2$, respectively. Both of the two processes converge to the stationary point.}
\end{figure}

Based on our model and simulation, we found that evolutionary stability in the eco-evolutionary game is guaranteed when the stochastic fluctuations are considered. In our model, the population size for each strategy is determined by the deterministic model and its variance is fixed and in directly proportional to the environmental carrying capacity for a large $\Omega$. The standard definition of evolutionary stability formulated for infinite populations \cite{hofbauer1998,nash1950,smith1982,lawlor1976} is a strong stability of the corresponding fixed point in the deterministic replicator equation, which is the theoretical concept- evolutionarily stable strategy (ESS). In the stochastic replicator dynamics with infinite populations subject to external noise conducted by Fudenberg and Harris, the population stays nearly all the time close to an ESS if the ESS corresponds to an interior state or if the payoff matrix satisfies a certain definiteness condition \cite{Fudenberg1992,traulsen2006}. We got a similar result in the eco-evolutionary game dynamics under environmental feedbacks, in which the population size is finite but not fixed. Therefore, stochastic fluctuations also won't impact disruptively on the dynamics of the eco-evolutionary game comparing to the deterministic equation \cite{chao2023}.

Our calculations of the fluctuation range ($\eta$ and $\xi$ in this study) are dependent on the assumption of a large environmental carrying capacity. The fluctuation is of the order $\Omega^{-\frac{1}{2}}$ relative to the macroscopic quantities and adds revisional terms to the macroscopic equations. The effects of higher order terms are neglected reasonably for large $\Omega$. Therefore, unlike the intrinsic noise of the evolutionary game dynamics, the fluctuation is related to the environmental carrying capacity instead of the population size in the eco-evolutionary game dynamics. However, when $\Omega$ is small, the linear noise approximation is disabled and thus Eqs.~(9) should be modified accordingly. It is thus no longer true that the fluctuation range is fixed even the variances of $\eta$ and $\xi$ are constants for a long time.

Additionally, we provide the following explanations: 1) Although the discussions in Sect. IV and V are based on the systems possessing one interior equilibrium, our derivation could be expanded to the general situation, i.e., these results will hold if the corresponding deterministic equations has more than one interior equilibria with asymptotically stable. 2) In our stochastic model, the environmental carrying capacity, which might be a finite-size constant in the deterministic model, is regarded as an mean value of the maximum population size. 3) Since we focus on the peak around the macroscopic values and give the power expansion by the $\Omega^{-1/2}$, we didn't consider the stochastic feature when $n_{1(2)}$ is close to the boundary for the the environmental carrying capacity, i.e., $p_1+p_2<<1$.

\section{Summary}
In this study, we focus on the properties of stochastic fluctuations and study the stochastic local stability in two-strategy eco-evolutionary game dynamics with environmental feedbacks. In our model, the population size is finite and varied. Through the $\Omega$-expansion method, the population dynamic is divided into two parts: the macroscopic equations and the moment equations. The macroscopic equations are equivalent to the deterministic model of eco-evolutionary game. By studying the moments equations, we discuss the stochastic local stability of the interior equilibrium in the macroscopic equation. It is shown that local stochastic fluctuations converge to a constant for a long time if the interior equilibrium is asymptotically stable. Therefore, the dynamical stability of the eco-evolutionary game, including the concept of eco-ESS proposed in our previous work, can be expanded to the stochastic case approximately.

Because we expand the equation in the power of the environmental carrying capacity $\Omega$, the results are affected by the assumption of a large environmental carrying capacity. With a decrease in environmental carrying capacity, this formalism is ineffective gradually. Higher order terms beyond the linear noise approximation should be considered for smaller environmental carrying capacity.

\section*{Acknowledgments}
This project was supported by the National Natural Science Foundation of China (no. 32171482).
\nocite{*}
\bibliographystyle{unsrt}

\appendix
\section{The expansion form of $\mathrm{e}^{1-(p_1+p_2)+e_1}$ and $\mathrm{e}^{1-(p_1+p_2)+e_2}$ in the power of $\Omega$}
\begin{eqnarray}
\mathrm{e}^{1-(p_1+p_2)+e_1}&=&\exp\left[ 1- \Omega^{-1} n_1-\Omega^{-1} n_2+ \Omega^{-1} n_1 a_{11} +\Omega^{-1} n_2 a_{12}\right]\nonumber\\[4pt]
&=&\exp \left[ 1+ (p_1+\Omega^{-1/2}\xi)(a_{11}-1)+(p_2+\Omega^{-1/2}\eta)(a_{12}-1)  \right]\nonumber\\[4pt]
&=&\mathrm{e}^{ 1+ p_1(a_{11}-1)+p_2(a_{12}-1)}\exp \left[\Omega^{-1/2}\xi(a_{11}-1)+\Omega^{-1/2}\eta(a_{12}-1)  \right]\nonumber\\[4pt]
&=&\mathrm{e}^{1+p_1(a_{11}-1)+p_2(a_{12}-1)}\times\nonumber\\[8pt]
&&\qquad\qquad\times\left[1+\Omega^{-\frac{1}{2}}[\xi(a_{11}-1)+\eta(a_{12}-1)]+\frac{1}{2}\Omega^{-1}[\xi(a_{11}-1)+\eta(a_{12}-1)]^2\right]\nonumber\\[8pt]
\end{eqnarray}

\begin{eqnarray}
\mathrm{e}^{1-(p_1+p_2)+e_2}&=&\exp\left[ 1- \Omega^{-1} n_1-\Omega^{-1} n_2+ \Omega^{-1} n_1 a_{21} +\Omega^{-1} n_2 a_{22}\right]\nonumber\\[4pt]
&=&\exp \left[1+ (p_1+\Omega^{-1/2}\xi)(a_{21}-1)+(p_2+\Omega^{-1/2}\eta)(a_{22}-1)  \right]\nonumber\\[4pt]
&=&\mathrm{e}^{1+ p_1(a_{21}-1)+p_2(a_{22}-1)}\exp \left[\Omega^{-1/2}\xi(a_{21}-1)+\Omega^{-1/2}\eta(a_{22}-1)  \right]\nonumber\\[4pt]
&=&\mathrm{e}^{1+p_1(a_{21}-1)+p_2(a_{22}-1)}\times\nonumber\\[8pt]
&&\qquad\qquad\times\left[1+\Omega^{-\frac{1}{2}}[\xi(a_{21}-1)+\eta(a_{22}-1)]+\frac{1}{2}\Omega^{-1}[\xi(a_{21}-1)+\eta(a_{22}-1)]^2\right]\nonumber\\[8pt]
\end{eqnarray}

\section{The solutions of  Eqs.~(21)}
According to the solution formula for the linear differential equations, the solutions to  Eqs.~(21) are
\begin{eqnarray}
 \left( \begin{array}{c}
     \langle\xi(t)\rangle \\
     \langle\eta(t)\rangle
  \end{array}\right)=\exp{\left[\mathbf{A}(t-t_0)\right]} \left( \begin{array}{c}
      \langle\xi(0)\rangle \\
  \langle\eta(0)\rangle
  \end{array}\right),
\end{eqnarray}
with
\begin{eqnarray}
\mathbf{A}&=&
\left( \begin{array}{cc}
      p_1^*\beta (a_{11}-1) & p_1^*\beta  (a_{12}-1)\\
      p_2^*\beta (a_{21}-1) & p_2^*\beta  (a_{22}-1)
  \end{array}\right)\nonumber\\[8pt]
  &=&\frac{\beta(\ln \beta-1)}{a_{11}a_{22}-a_{12}a_{21}-a_{11}-a_{22}+a_{12}+a_{21}}\left( \begin{array}{cc}
      (a_{11}-1)(a_{12}-a_{22}) & (a_{12}-1)(a_{12}-a_{22})\\
      (a_{21}-1)(a_{21}-a_{11}) & (a_{22}-1)(a_{21}-a_{11})
  \end{array}\right)\nonumber\\
\end{eqnarray}
being the corresponding coefficient matrix in Eqs.~(21) and $\exp{\left[\mathbf{A}(t-t_0)\right]}=(\mathrm{e}^{\lambda_1 t}V_1,\,\mathrm{e}^{\lambda_2 t}V_2)$, where $\lambda_{1(2)}$ are the eigenvalues of matrix $\mathbf{A}$ and $V_{1(2)}=(v_{1(2)},\,1)^T$ are the corresponding eigenvectors with
\begin{eqnarray}
\lambda_{1(2)}&=&\frac{\beta(\ln \beta-1)}{a_{11}a_{22}-a_{12}a_{21}-a_{11}-a_{22}+a_{12}+a_{21}} \left[(a_{11}-1)(a_{12}-a_{22})+(a_{22}-1)(a_{21}-a_{11})\pm\sqrt{\Delta} \right] \nonumber \\[8pt]
\end{eqnarray}
and
\begin{eqnarray}
v_{1(2)}&=&\frac{\beta(\ln \beta-1)}{a_{11}a_{22}-a_{12}a_{21}-a_{11}-a_{22}+a_{12}+a_{21}}  \times \nonumber\\[8pt]
&&\qquad \times \left[ (a_{11}-1)(a_{12}-a_{22})-(a_{22}-1)(a_{21}-a_{11})\pm\frac{\sqrt{\Delta}}{2(a_{21}-1)(a_{21}-a_{11})}\right],  \nonumber\\[8pt]
\end{eqnarray}
with $\Delta=[(a_{11}-1)(a_{12}-a_{22})+(a_{22}-1)(a_{21}-a_{11})]^2-4[(a_{11}-1)(a_{12}-a_{22})(a_{22}-1)(a_{21}-a_{11})-(a_{12}-1)(a_{12}-a_{22})(a_{21}-1)(a_{21}-a_{11})]$.
Therefore,
\begin{eqnarray}
     \langle\xi(t)\rangle&=& \mathrm{e}^{\lambda_1 t}v_1 \langle\xi(0)\rangle+\mathrm{e}^{\lambda_2 t}v_2 \langle \eta(0)\rangle, \nonumber\\ [4pt]
     \langle\eta(t)\rangle&=& \mathrm{e}^{\lambda_1 t}\langle\xi(0)\rangle+\mathrm{e}^{\lambda_2 t} \langle \eta(0)\rangle.
\end{eqnarray}

\section{The solutions of  Eqs.~(22)}
For the non-homogeneous linear differential equations (22), the solutions are
\begin{eqnarray}
 \left( \begin{array}{c}
     \langle\xi(t)^2\rangle^s \\
 \langle\xi(t)\eta(t)\rangle^s \\
     \langle\xi(t)^2\rangle^s
  \end{array}\right)&=&\mathrm{e}^{\mathbf{A^\prime}t}  \left( \begin{array}{c}
     \langle\xi(0)^2\rangle^s \\
 \langle\xi(0)\eta(0)\rangle^s \\
     \langle\xi(0)^2\rangle^s
  \end{array}\right)+\mathrm{e}^{\mathbf{A^\prime}t} \int^t_0\exp{\left[-\mathbf{A^\prime}s \right]} \left( \begin{array}{c}
     \langle\xi(s)^2\rangle^s \\
           0                   \\
     \langle\xi(s)^2\rangle^s
  \end{array}\right)\mathrm{d}s,
\end{eqnarray}
with $\mathbf{A^\prime}$ being the coefficient matrix of Eqs.~(22)
\begin{eqnarray}
  \left ( \begin{array}{l ll}
  2\beta p_1^*  (a_{11}-1)&\qquad 2\beta p_1^* (a_{12}-1)                                           &\qquad 0  \\
  2 \beta p_2^* (a_{21}-1)            &\qquad \beta  p_1^*  (a_{11}-1)+ \beta  p_2^* (a_{22}-1)                 &\qquad2\beta p_1^* (a_{12}-1) \\
          0                      &\qquad 2\beta p_2^* (a_{21}-1)                                             &\qquad 2\beta p_2^*(a_{22}-1)
  \end{array}\right).
\end{eqnarray}
and $\mathrm{e}^{\mathbf{A^\prime}t}=(e^{-\lambda_1 t} u_1,\,e^{-\lambda_2 t} u_2,\,e^{-\lambda_3 t} u_3)$,
where $\lambda_i$ are the eigenvalues of $\mathbf{A^\prime}$ and $u_i$ are the corresponding eigenvectors. Note that $\lambda_{1,\,2}$ are equal to the matrix $\mathbf{A}$ of Appendix B and $\lambda_3=2\beta \left [   p_1^*  (a_{11}-1)+   p_2^*(a_{22}-1) \right ]$, for which the proof is provided in the reference \cite{yi2007}.

\end{document}